\documentclass[aps,prl,twocolumn,showpacs]{revtex4-1}%
\usepackage{amsfonts}
\usepackage{amsmath}
\usepackage{amssymb}
\usepackage{graphicx}
\usepackage{dcolumn}
\usepackage{bm}

\begin{document}

\title{Field dependence of the vortex core size probed by STM}
 
\author{A. Fente}
\affiliation{Laboratorio de Bajas Temperaturas, Departamento de F\'isica de la Materia
Condensada, Instituto de Ciencia de Materiales Nicol\'as Cabrera and
Condensed Matter Physics Center (IFIMAC), Universidad Aut\'onoma de Madrid,
Spain}
\affiliation{Unidad Asociada de Bajas Temperaturas y Altos Campos Magn\'eticos, UAM, CSIC, Spain}
\author{E. Herrera}
\affiliation{Laboratorio de Bajas Temperaturas, Departamento de F\'isica de la Materia
Condensada, Instituto de Ciencia de Materiales Nicol\'as Cabrera and
Condensed Matter Physics Center (IFIMAC), Universidad Aut\'onoma de Madrid,
Spain}
\affiliation{Unidad Asociada de Bajas Temperaturas y Altos Campos Magn\'eticos, UAM, CSIC, Spain}
\author{I. Guillam\'on}
\affiliation{Laboratorio de Bajas Temperaturas, Departamento de F\'isica de la Materia
Condensada, Instituto de Ciencia de Materiales Nicol\'as Cabrera and
Condensed Matter Physics Center (IFIMAC), Universidad Aut\'onoma de Madrid,
Spain}
\affiliation{Unidad Asociada de Bajas Temperaturas y Altos Campos Magn\'eticos, UAM, CSIC, Spain}
\author{H. Suderow}
\affiliation{Laboratorio de Bajas Temperaturas, Departamento de F\'isica de la Materia
Condensada, Instituto de Ciencia de Materiales Nicol\'as Cabrera and
Condensed Matter Physics Center (IFIMAC), Universidad Aut\'onoma de Madrid,
Spain}
\affiliation{Unidad Asociada de Bajas Temperaturas y Altos Campos Magn\'eticos, UAM, CSIC, Spain}
\author{S. Ma\~nas-Valero} 
\affiliation{Instituto de Ciencia Molecular (ICMol), Universidad de Valencia, Catedr\'atico Jos\'e Beltr\'an 2, 46980 Paterna, Spain}
\author{M. Galbiati} 
\affiliation{Instituto de Ciencia Molecular (ICMol), Universidad de Valencia, Catedr\'atico Jos\'e Beltr\'an 2, 46980 Paterna, Spain}
\author{E. Coronado} 
\affiliation{Instituto de Ciencia Molecular (ICMol), Universidad de Valencia, Catedr\'atico Jos\'e Beltr\'an 2, 46980 Paterna, Spain}
\author{V. G. Kogan}
\affiliation{Ames Laboratory and Department of Physics \& Astronomy, 
Iowa State University, Ames, Iowa 50011, USA}

\date{\today}

 \begin{abstract}
We study the spatial distribution of the density of states (DOS) at zero bias $N(\bm r)$ in the mixed state of single and multigap superconductors. We provide an analytic expression for $N(\bm r)$ based on deGennes' relation between DOS and the order parameter that reproduces well Scanning Tunneling Microscopy (STM) data in several superconducting materials. In the single gap superconductor $\beta$-Bi$_2$Pd, we find that $N(\bm r)$ is governed by a length scale $\xi_H=\sqrt{\phi_0/2\pi H}$, which decreases in rising  fields. The vortex core size $\cal C$, defined via the slope of the order parameter at the vortex center, $\cal C$ $\propto (d \Delta /dr |_{r \to 0})^{-1}$, differs from $\xi_H$ by a material dependent numerical factor. For two gap superconductors 2H-NbSe$_{1.8}$S$_{0.2}$ and 2H-NbS$_2$, we find that $\cal C$ is field independent and has the same value for both bands. We conclude that, independently of the magnetic field induced variation of the order parameter values in both bands, the spatial variation of the order parameter close to the vortex core is the same for all bands.
\end{abstract}
 \maketitle

The spatial distribution of the quasiparticles density of states (DOS) within the vortex lattice (VL) is intimately related to the spatial distribution of the order parameter. The latter is governed by the coherence length $\xi$ which sets the size of the vortex core and by the applied magnetic field which fixes the intervortex spacing. 

There are  a few definitions of $\xi$ used in literature,  adjusted to a particular problem at hand, see e.g. Ref.\,\onlinecite{KZ}. Within  this work, $\xi$ is associated with the vortex core size $\cal C$, which is related to the order parameter slope at the vortex center, $d\Delta/dr|_{r\to 0}\propto 1/\xi\propto 1/$$\cal C$. It was suggested theoretically that $\cal C$ shrinks with the increasing magnetic field $H$ \cite{KZ}. In fact, interpreting  $\mu$SR data on  various  materials, it was deduced that $\xi$ decreases with increasing fields \cite{Sonier}. This conclusion was obtained with the help of London-based models for the field distribution within VL. In these models, $\xi$ enters as a cutoff, which restricts their applicability. For this reason, extracting $\xi(H)$ from $\mu$SR data can hardly be considered as direct. Similar shortcomings can be attributed to $\xi(H)$  deduced  from the  magnetization data \cite{M(H)}.

The Scanning Tunneling Microscopy (STM) has the advantage of directly probing the spatial distribution of the quasiparticles DOS within the vortex lattice. The DOS depends on the  value of the order parameter $\Delta(\bm r)$ and can be used, in principle, to map $|\Delta(\bm r)|$ within VL.  This was done within the microscopic quasi-classical formalism  by U. Klein \cite{Klein} for clean Nb and later by N. Nakai {\it et al} for 2H-NbSe$_2$ \cite{Machida}. Similar approaches have been  applied to other materials, such as nickel-borocarbides and pnictide compounds  \cite{ReviewHirschfeld,ReviewFisher,ReviewSUST}, requiring always a detailed knowledge of the normal phase properties \cite{Hess89,Hess90,ReviewFisher,ReviewSUST}. However, extracting a value for $\xi$ or obtaining order parameter variations in different bands from STM data remains highly non-trivial. There is thus a need to discuss within a simple model the spatial distribution of the DOS within the vortex lattice unit cell.

Perhaps the most compact and  simple result for the DOS distribution in the mixed state was given by P.G. deGennes in the work on  dirty superconductors \cite{DeGennes1,Maki}.  Following this work, we offer here a phenomenological scheme to describe  the STM data on zero-energy DOS for materials with hexagonal vortex lattices. 
If needed, the approach can be generalized for other VL symmetries and for anisotropic superconductors. We show that the DOS distribution within VL can be well described by the model for a single- and   two-gap  superconductors. For the single gap case we find that the core size $\cal C$ is  proportional to a universal length $\xi_H=  \sqrt{\phi_0/2\pi H}$. When $H$ approaches the upper critical field $H_{c2}$, $\xi_H $ coincides with the commonly used coherence length $\xi_{c2}=  \sqrt{\phi_0/2\pi  H_{c2}}$. This behavior agrees with predictions \cite{KZ} for clean materials. For the two-gap samples, we find nearly field independent $\cal C$ and equal core sizes, ${\cal C}_{1}={\cal C}_2$.

We have chosen $\beta$-Bi$_2$Pd for a single gap superconductor, and 2H-NbSe$_{1.8}$S$_{0.2}$ and 2H-NbS$_2$ which are isotropic multigap superconductors. The tunneling conductance and VL in 2H-NbSe$_{1.8}$S$_{0.2}$ is first reported here whereas the data for $\beta-$Bi$_2$Pd and 2H-NbS$_2$ are taken from our previous work \cite{Bi2Pd,NbS2}. Details of the sample preparation and of DOS measurements are given in the supplemental material \cite{Suppl}. Superconducting parameters of the compounds are given in Table I, and their zero field tunneling conductance curves are shown in Fig.\,1. Note that the critical temperatures of the  three compounds are  similar, although $\xi_{c2}$ obtained from the upper critical fields vary by a factor of three. Mean free paths have been estimated from resistivity measurements, yielding values slightly above or comparable to the coherence lengths. $\beta-$Bi$_2$Pd is clearly a single gap superconductor ($\Delta=0.76$meV, Fig.\,1) with a zero field  conductance following s-wave BCS theory and shows a hexagonal VL \cite{Bi2Pd}. The zero field  conductance of both 2H-NbSe$_{1.8}$S$_{0.2}$ and 2H-NbS$_2$ can be fitted using BCS theory (red lines in Fig.\,1) with two  gaps (see table I)  \cite{Hess89,Hess90,Guillamon08}.

 \begin{figure}[htb]
\begin{center}
\includegraphics[width=8.cm] {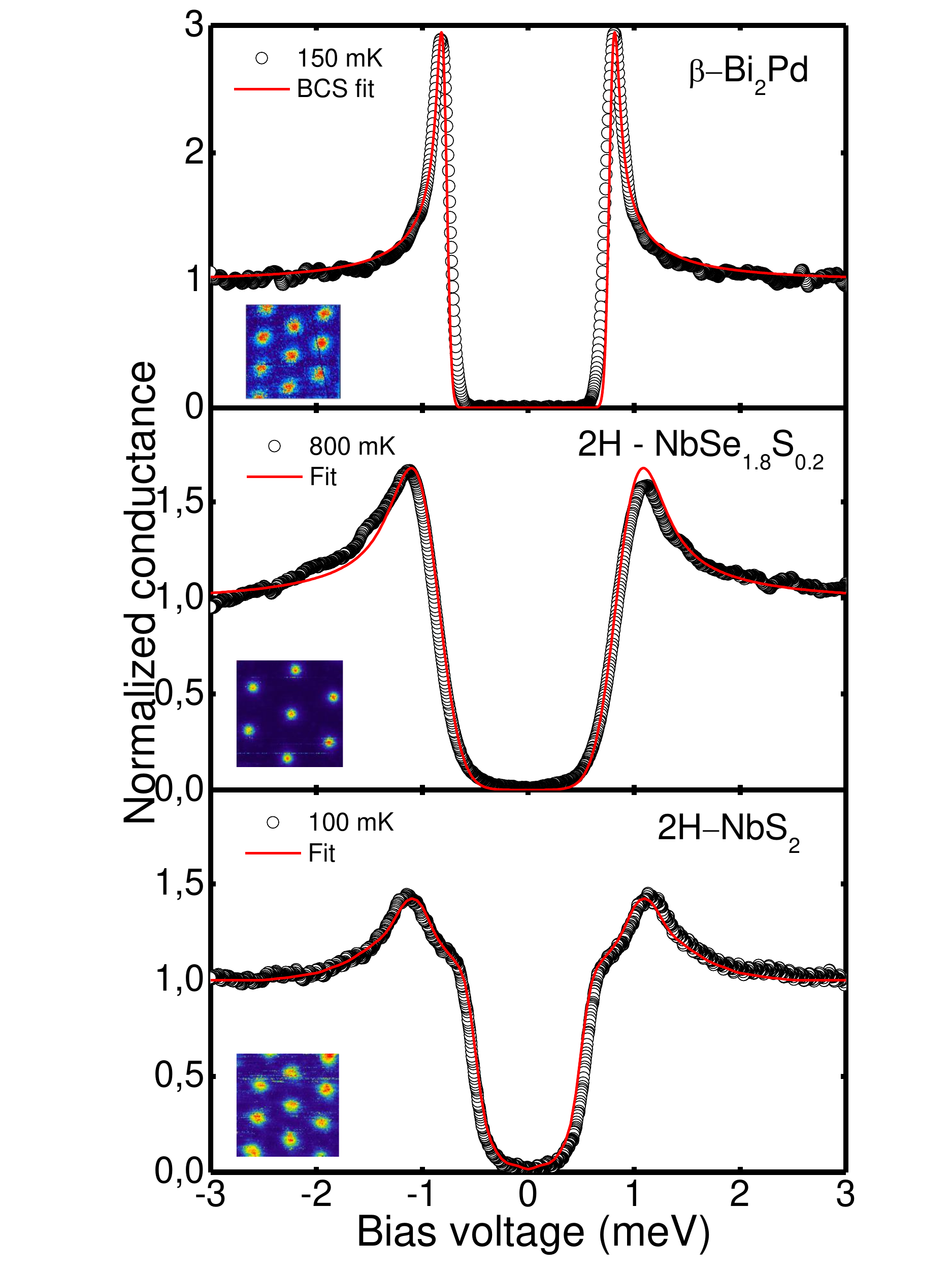}
\vskip -.6 cm
\caption{Zero field tunneling conductance curves for $\beta$-Bi$_2$Pd (upper panel), 2H-NbSe$_{1.8}$S$_{0.2}$ (middle panel) and 2H-NbS$_2$ (lower panel). Fits to BCS theory are given by red lines (see \protect\cite{Suppl}). Temperatures at which the data were taken are given in each panel. Insets show VL images (lateral sizes of 450 nm, 360 nm and 290 nm, respectively) obtained from the zero bias conductance  at, from top to bottom, 0.05 T, 0.1 T, and 0.15 T.}
\label{f1}
\end{center}
\end{figure}

\begin{table}[h]
\begin{tabular}{c | c | c | c | c}
Compound & $T_c$ (K) & $H_{c2}$ (T) & $\xi_{c2}$ (nm) & $\Delta$ (meV)\\
\hline
$\beta$-Bi$_2$Pd & 5 & 0.6 & 23 & 0.8\\
2H-NbSe$_{1.8}$S$_{0.2}$ & 7  & 7  & 7  & 0.8; 1 \\
2H-NbS$_2$ & 5.7  & 2.5  & 12 & 0.5; 1 \\
\end{tabular}
\caption{Superconducting parameters of the compounds studied.  The gap values are obtained from the fits (red lines in Fig.\,1), see also supplemental material.}
\end{table}

To obtain tunneling conductance vs distance from the vortex center, we select single vortices out of zero bias conductance images and evaluate angular  averages of the  normalized conductance $\sigma_0$ for each $r$. We define $\sigma$ as:
\begin{equation}
\sigma =\frac{\sigma_0(r)-\sigma_0(r^*)}{\sigma_0(0)-\sigma_0(r^*)}\,.
\label{eq1}
\end{equation}
where $r^*$ is the distance from the vortex center to the point where the tunneling conductance is minimum (in the hexagonal lattice, the center of an equilateral vortex triangle). 

The zero-bias DOS $N(\bm r)$ in large fields of the mixed state at low temperatures and in the dirty limit was given by P.G. deGennes  \cite{DeGennes1,Saint-James}:
 \begin{equation}
\frac{N(\bm r)}{N_n} = 1-\frac{|\Delta(\bm r)|^2}{\Delta_0^2 }\,,  
\label{eq0}
\end{equation}
$N_n$ is DOS in the normal phase.
When  $\Delta \to  0$, as e.g. at $H_{c2}$ or at  vortex centers, $N\to N_n$, as it should. This remarkable relation expresses the local DOS in terms of the order parameter at the same point. 
Precise value of the constant $\Delta_0$ (on the order of zero-$T$  BCS gap) will not affect our analysis. Note that this relation does not account for possible core states \cite{DeGennes-Caroli} (see supplemental material).

One can argue that  $N(\bm r)$  depends only on even powers of $\Delta$. Within the   Eilenberger version of the BCS theory \cite{E} the superconductivity is described by   Gor'kov Green's functions integrated over energy, $f, f^+$ and $g$, which depend on Matsubara frequencies $\omega$ and are related by $g^2=1-ff^+$.  The DOS  as a function of energy $\epsilon$ is given by  $N=N_n {\rm Re}[g(\omega\to i \epsilon)]$, where  $f$ and $f^+$ are $\propto |\Delta |$.   Hence, $g$ depends only on $|\Delta|^2$ and so does $N$. 

In  large fields of the mixed state the order parameter is suppressed relative to the zero-field value $\Delta_0$. The ratio $\Delta^2(\bm r) /\Delta_0^2  $ is small and terms correcting  Eq.\,(\ref{eq0}) of the order $\Delta^4(\bm r) /\Delta_0^4 $ are  smaller yet. Hence, Eq.\,(\ref{eq0}) is likely to hold  not only in the dirty limit  \cite{remark} and we take it  is a basis of our phenomenological model.  

The order parameter for a single vortex in isotropic superconductors can be  approximated by  \cite{Schmid,Clem} :
 \begin{equation}
\frac{\Delta(r)}{\Delta_0(T)}= \frac{r}{\sqrt{r^2+{\cal C}^2}} \,,\label{eq1}
\end{equation}
where the core size ${\cal C}$   is  of the order of $\xi$. 
 This function reproduces the expected behavior for $r\to 0$, where $d\Delta/dr|_{r\to 0}=\Delta_0/{\cal C}$, for $r\gg{\cal C}$, $\Delta\to\Delta_0 $, the order parameter of uniform zero-field state. Minimizing the  Ginzburg-Landau energy functional, Z. Hao and J. Clem deduced ${\cal C}=\xi \sqrt{2}$ to fit magnetization data in large fields  \cite{Hao-Clem}.
 
For the hexagonal VL we  use Wigner-Zeitz approximation and  consider the unit cell as a circle of a radius $a$ such that $\pi a^2=\phi_0/B$, $\phi_0$ is the flux quantum and $B$ is the magnetic induction. The cell radius  $a$ is close to the half of intervortex distance $L$:  $2a/L=\sqrt{2\sqrt{3}/\pi}\approx 1.05$. 
 
The derivative $d\Delta/dr$ should vanish at the cell boundary $r=a$. The function (\ref{eq1}) does not satisfy this condition, although for $a\gg {\cal C}$ this derivative is small. To correct this, we modify the function to
 \begin{equation}
\frac{\Delta(r)}{\Delta_0(B,T)} =\frac{r}{\sqrt{r^2+ {\cal C}^2}} \exp\left[-\frac{ r^2 {\cal C}^2}{ 2a^2( {\cal C}^2+a^2)}\right]\,
\label{eq2}
\end{equation}
which satisfies $d\Delta/dr=0$ at $r=a$ and for $a\gg{\cal C}$ approaches the maximum at the cell boundary exponentially slow. Expressions (\ref{eq1}) and (\ref{eq2})   practically coincide within the core $r< {\cal C}$, for  larger $r$ the new function varies slower than for a single vortex. The slope $d\Delta/dr|_{r\to 0}=\Delta_0/{\cal C}$, so that $\cal C$ can still be taken as the core size. 
 
Next, we observe that the normalizing constant $\Delta_0(B,T)$ drops off the measured quantity  
\begin{equation}
\sigma=\frac{N(r)-N(a)}{N(0)-N(a)}= \frac{\Delta^2(a)-\Delta^2(r)}{\Delta^2(a)-\Delta^2(0)}=1- \frac{ \Delta^2(r)}{\Delta^2(a) }\, 
\label{data}
\end{equation}
since  $\Delta(0)=0$ and $\Delta(a)=\Delta_0(B)$. 
We now take $a$ as a unit length to obtain:
\begin{eqnarray}
\sigma=1&-& \frac{ \rho^2(1+\eta^2)}{\rho^2+\eta^2}\,\exp \frac{\eta^2(1-\rho^2)}{1+\eta^2}\,,\nonumber\\
 \rho&=& r/a \,, \qquad \eta= {\cal C}/a \,.\qquad
\label{sigma}
\end{eqnarray}

\begin{figure}[htb]
\begin{center}
 \includegraphics[width=8.cm] {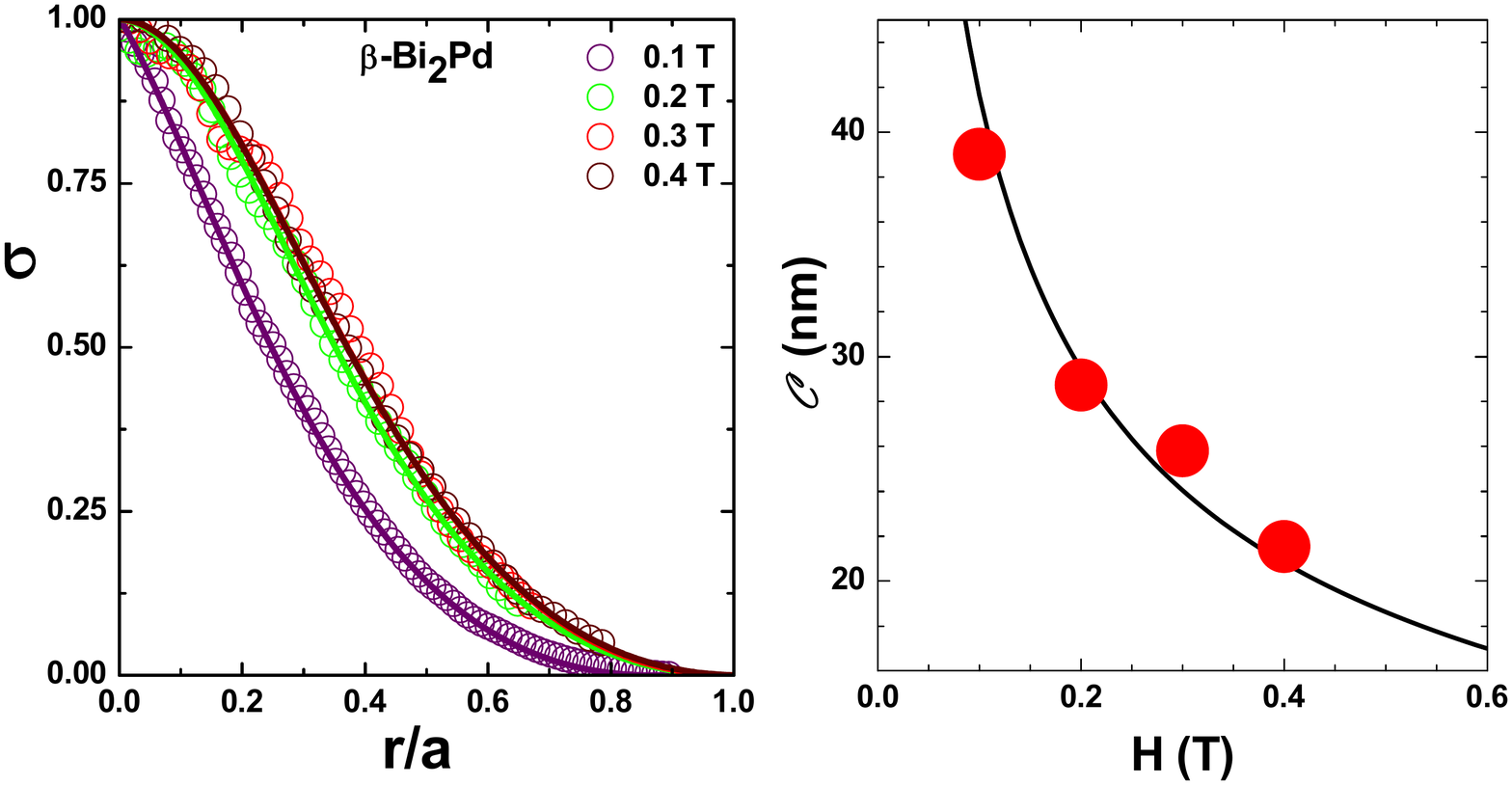}
\vskip -1 cm
\caption{The left panel:  the tunneling conductance $\sigma$ of Eq.\,(\ref{sigma}) vs distance $r$ from the vortex center, normalized to the cell radius $a$, for $\beta$-Bi$_2$Pd and in    fields indicated. Data are taken at 0.15\,K and have been obtained from images of vortices radially averaged and normalized as described in the text. The right panel:  the core size $\cal C$ of Eq.\,(\ref{eq7}) vs $H$. Dots are the values of ${\cal C} =\eta\, a$ obtained from the fits of the left panel   with $a$ being  the   Wigner-Seitz cell size. The line is ${\cal C}$ calculated with $\eta\approx 0.5$ found in the fits.}
\label{f2}
\end{center}
\end{figure}

Let us first focus on $\beta$-Bi$_2$Pd (Fig.\,2). Fitting the data to 
Eq.\,(\ref{sigma}) and treating $\eta$ as a fit-parameter we can extract  the core size ${\cal C}$. The good quality of the fit validates   our model   as able to provide quantitative description of the STM data. The fits yield values of $\eta$ around $0.50 \pm 0.08$ in a field range where $H$ changes by a factor of 4. Since $\eta\approx\,\,$const, we have 
\begin{equation}
 {\cal C}=\eta a = \eta\sqrt{\frac{\phi_0}{\pi H}} \,.
\label{eq7}
\end{equation}
Hence, the core size ${\cal C}$ varies with applied field as $1/\sqrt{H}$, the dependence   deduced from the $\mu$SR data on many materials  \cite{Sonier}. This is highlighted by the red points in the right panel of Fig.2, which provide ${\cal C}$ vs magnetic field.

The theory of Ref.\,\onlinecite{KZ} suggests that  in large fields the coherence length as a function of field should behave as $\xi\approx A/\sqrt{H}$ (except in the extreme dirty limit or at high  temperatures). In particular, this relation should hold at the upper critical field  $H_{c2}$. This   gives the constant $A=\sqrt{\phi_0/2\pi}$ so that we have
\begin{equation}
 \xi =\sqrt{\frac{\phi_0}{2\pi H}} \equiv \xi_H\,.
\label{eq8}
\end{equation}
The core size, therefore, is
\begin{equation}
 {\cal C}=\eta a = \eta\sqrt{\frac{\phi_0}{\pi H}}=\eta\sqrt{2}\,\xi_H\,.   
 \label{eq9}
\end{equation}
Using $\eta\approx 0.5$ obtained by the fits,  we find ${\cal C}\approx 0.7\, \xi_H$.
Eq.\,(\ref{eq9}) implies that the order parameter and DOS distributions within VL in large fields of one-band isotropic materials are governed by a {\it universal} length $\xi_H$ of Eq.\,(\ref{eq8}). As $H\to H_{c2}$, $\xi_H$ reaches the value of the standard coherence length $ \xi_{c2}=\sqrt{\phi_0/2\pi H_{c2}}$.  Our results suggest that by a proper rescaling,  distributions of zero-bias DOS  in large-field  vortex lattices in one-band isotropic superconductors can be reduced to a nearly universal form. 
 
We now turn to multigap superconductors 2H-NbSe$_{1.8}$S$ _{0.2}$ and 2H-NbS$_2$. To generalize our model to the two-gap situation we assume that the order parameter takes values $\Delta_1$ and $\Delta_2$ on two bands and write the spatially dependent density of states as
\begin{equation}
\frac{N(\bm r)}{N_n}=1- n_1\frac{\Delta_1^2(\bm r)}{\Delta_{10}^2}- n_2\frac{\Delta_2^2(\bm r)}{\Delta_{20}^2}
\label{equ_10}
\end{equation}
where  $n_{1,2}$ are partial DOS in the normal state, $n_{1}+n_2=1$. 
Each $\Delta_\nu$ ($\nu=1,2$) satisfies the boundary condition $d\Delta_\nu/dr=0$ at the cell boundary $r=a$ because each one should be periodic in the vortex lattice: 
 \begin{equation}
\frac{\Delta_\nu(\bm r)}{\Delta_{ 0\nu} } =\delta_\nu=\frac{\rho}{\sqrt{\rho^2+ \eta_{\nu}^2}} \exp\left[-\frac{\rho^2 \eta_{\nu}^2}{ 2 ( \eta_{\nu}^2+1)}\right],
\label{A2}
\end{equation}
$\eta_\nu={\cal C}_{\nu}/a$, $\nu=1,2$.   Substituting this in Eq.\,(\ref{equ_10}), we obtain:
 \begin{equation}
\sigma=1- \frac{\delta_1^2( \rho) +\gamma\, \delta_2^2(\rho)}{\delta_1^2(1) +\gamma \,\delta_2^2(1) } \,, \quad \gamma=\frac{n_2}{n_1} .
\label{A3}
\end{equation}
Note that $\delta_{\nu}$ are normalized to corresponding $\Delta_{ 0\nu}$.

\begin{figure}[htb]
\begin{center}
\includegraphics[width=8.cm] {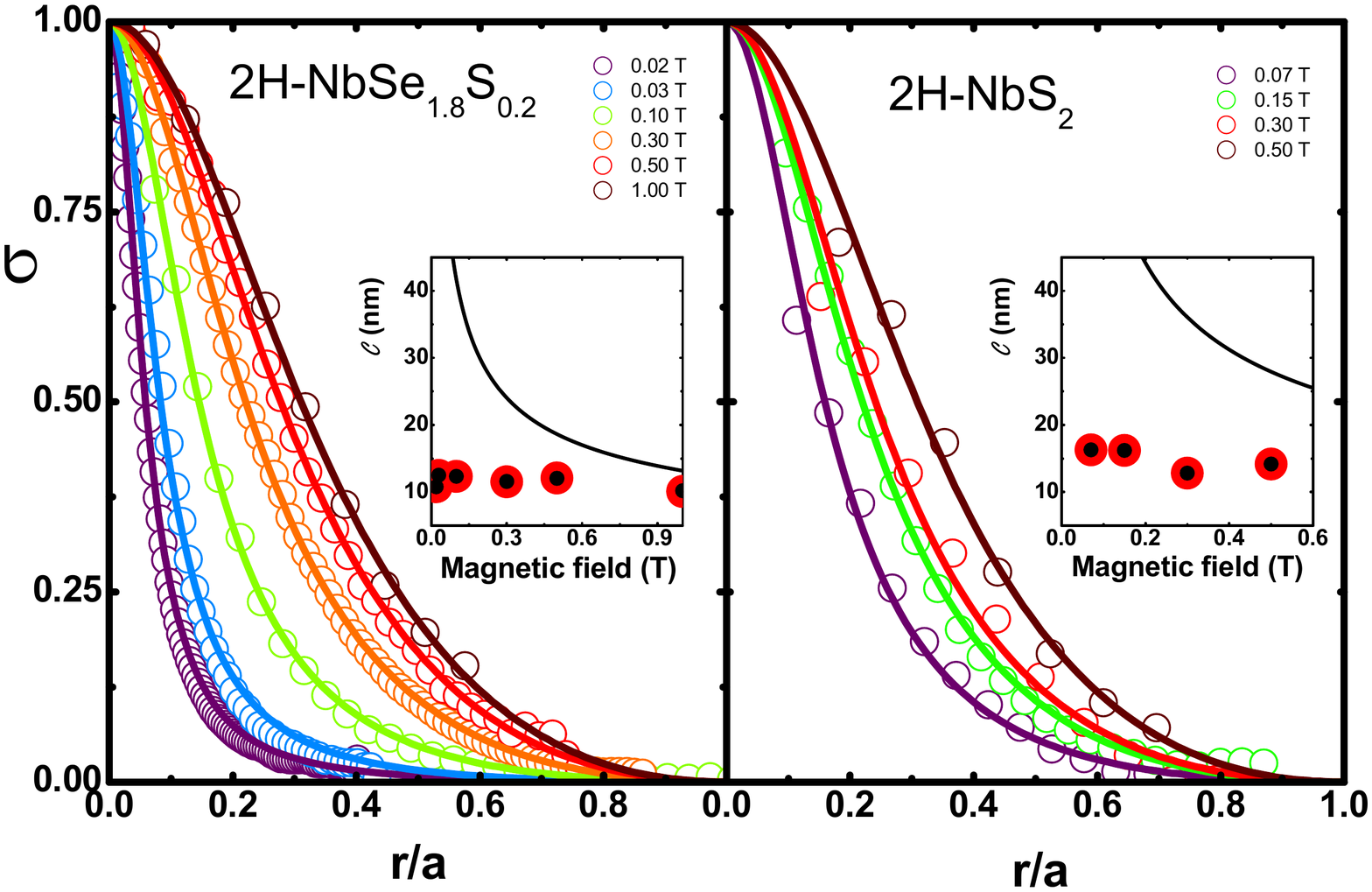}
\vskip -.5 cm
\caption{Normalized  conductance $\sigma$  for 2H-NbSe$_{1.8}$S$_{0.2}$ (left panel) and  for 2H-NbS$_2$ (right panel). Lines are fits as described in the text. Insets give the core size vs magnetic field ${\cal C}_{1,2}$ (red and black points respectively) 
and the line is ${\cal C}$ calculated with $\eta\approx 0.5$ for comparison.}
\label{f3}
\end{center}
\end{figure}

Fitting the data for $\sigma(r)$ we extract $\eta_1$ and $\eta_2$. Fig.\,\ref{f3} shows results of such a fitting. The good quality of the fits is remarkable.  Thus, the expression (\ref{A3}) describes well the spatial distribution of the DOS in two-gap systems. Using Eq.\,(\ref{equ_10}), we calculate the core sizes from the fitting parameters and find---within accuracy of our procedure---nearly equal and field independent core sizes ${\cal C}_1$ and ${\cal C}_2$. The spatial dependence of the conductance curves in 2H-NbSe$_{1.8}$S$_{0.2}$ and in 2H-NbS$_{2}$, plotted vs $r/a$ is clearly $H$ dependent (Fig.\,3). The density of states spreads considerably when applying a magnetic field, i.e. the parameters $\eta_\nu$ increase with the magnetic field. This means that ${\cal C}=\eta\, a$ no longer shrinks with the increasing vortex density, in fact the fits show that $\cal C$ are field independent. What is more, the slopes of the order parameter values in each band, ${\cal C}_1$ and ${\cal C}_2$, remains the same.
 
Let us discuss the magnetic field independence of $\cal C$. This is expected for superconductors in the dirty limit \cite{KZ}. In 2H-NbSe$_{1.8}$S$ _{0.2}$, there is a remarkable increase of H$_{c2}$ by a factor of two with respect to pure 2H-NbSe$_2$, and the zero bias peak at the vortex core is considerably suppressed (see \cite{Suppl}). The Fermi velocities range from $10^5\,$m/s to well above $10^6\,$m/s, leading to BCS zero-temperature coherence length values that range between 10 nm and 50 nm, in any case above the values $\xi_{c2}\approx 7$ nm obtained from H$_{c2}$ (see Table I) \cite{Mazin,Bandstructure}. This suggests that the influence of scattering is strong in 2H-NbSe$_{1.8}$S$ _{0.2}$.

In 2H-NbS$_2$, the residual resistivity shows the lowest values among the compounds discussed here (see \cite{Suppl}) and there is a clear zero bias peak at the vortex core, comparable to the peak observed in pure 2H-NbSe$_2$. The observed magnetic field independence of $\cal C$ is therefore unexpected in this material and requires more careful band dependent calculations. Probably, the influence of scattering is band dependent, giving different sensitivities to scattering on the zero bias peak and the spatial dependence of the order parameter values close to the vortex core ${\cal C}_{1,2}$. 

On the other hand, our model applied to  STM data on two-gap 2H-NbSe$_{1.8}$S$_{0.2}$ and 2H-NbS$_{2}$ shows that the length scales on which the order parameter changes in the two bands are, in fact, the same. This outcome is unexpected, one would think that at low temperatures the length scales at which order parameters change should be close to the BCS coherence lengths of the two bands, $\xi_{0\nu}\propto v_\nu/T_c$.  However, it has been shown time ago by B. Geilikman, R. Zaitsev, and
V. Kresin \cite{Kresin} and ``rediscovered" recently \cite{Jani,Kog-Schm} that near $T_c$ the two-gap Ginzburg-Landau theory yields the same length scales for variation of both $\Delta_1$ and $\Delta_2$. Our statement that the same  is true for low temperatures is, of course, 
based on phenomenological model  and as such might be corrected by a future self-consistent microscopic theory.
Our model, however, should not be far from reality, since we are able to reproduce low temperature STM data quite well.

To summarize, we argue that deGennes formula for the zero-bias DOS $N(\bm r)$ in the mixed state \cite{DeGennes1} can be used out of the dirty limit. Combining this  with the  Wigner-Zeitz  approximation for the order parameter within the VL unit cell and the known approximation for the order parameter distribution, we are able to reproduce the experimental $N(\bm r)$ for one- and two-gap materials. This allows us to extract the vortex core size $\cal C$ not as an imaginary boundary separating the ``normal" core from superconducting environment, but as a parameter characterizing the DOS  distribution in the vortex lattice. We find that, as predicted in \cite{KZ}, the core size shrinks as $1/\sqrt{H}$ in $\beta$-Bi$_2$Pd, whereas it remains magnetic field independent in multigap materials 2H-NbSe$_{1.8}$S$_{0.2}$ and in 2H-NbS$_2$. Furthermore, in the latter two compounds, there is no difference between the magnetic field dependence of the core size in both bands. $\beta$-Bi$_2$Pd is an isotropic material with a relatively large coherence length and behaves according to the expectations of \cite{KZ} for the clean limit. In 2H-NbSe$_{1.8}$S$_{0.2}$, scattering leads to the magnetic field independent core size expected in \cite{KZ} for the dirty limit. The result in 2H-NbS$_2$ suggests that the band dependence of electronic scattering is important to understand details of the density of states of multigap superconductors.

\section{Acknowledgements}

We would like to acknowledge Paul C. Canfield for fostering discussions between Ames and Madrid and for convincingly sharing his view about the relevance of high quality single crystal growth, and S. Vieira for discussions. The work was supported by the Spanish Ministry of Economy and Competitiveness (FIS2014-54498-R, MAT2014-56143-R, MDM-2014-0377 and MDM2015-0538, Network of Excellence in Molecular Nanoscience MAT2014-52919-REDC), by the Comunidad de Madrid through program Nanofrontmag-CM (S2013/MIT-2850), the Generalidad Valenciana through program Prometeo and by EU (Cost MP-1201 and COST CA-15128). E.H. acknowledges support of COLCIENCIAS Programa Doctorados en el Exterior Convocatoria 568-2012 and S.M. of MECD: FPU14/04407. M.G. acknowledges the European Union Horizon 2020 Marie Curie Actions under the project SPIN2D (H2020/2014-659378). We acknowledge SEGAINVEX workshop of UAM and Banco Santander. The work of I.G. receives support from Axa Research Fund, FP7-PEOPLE-2013-CIG 618321 and the European Research Council (grant agreement 679080). The work of V.K. was supported by the U.S. Department of Energy, Office of Basic Energy Sciences, Division of Materials Sciences and Engineering under contract No. DE-AC02-07CH11358.

\email[Corresponding authors: {kogan@ameslab.gov and hermann.suderow@uam.es}

\section{Supplemental material}
\subsection{Vortex lattice of 2H-NbSe$_{1.8}$S$_{0.2}$}

\begin{figure}[h]
\begin{center}
\includegraphics[width=9cm]{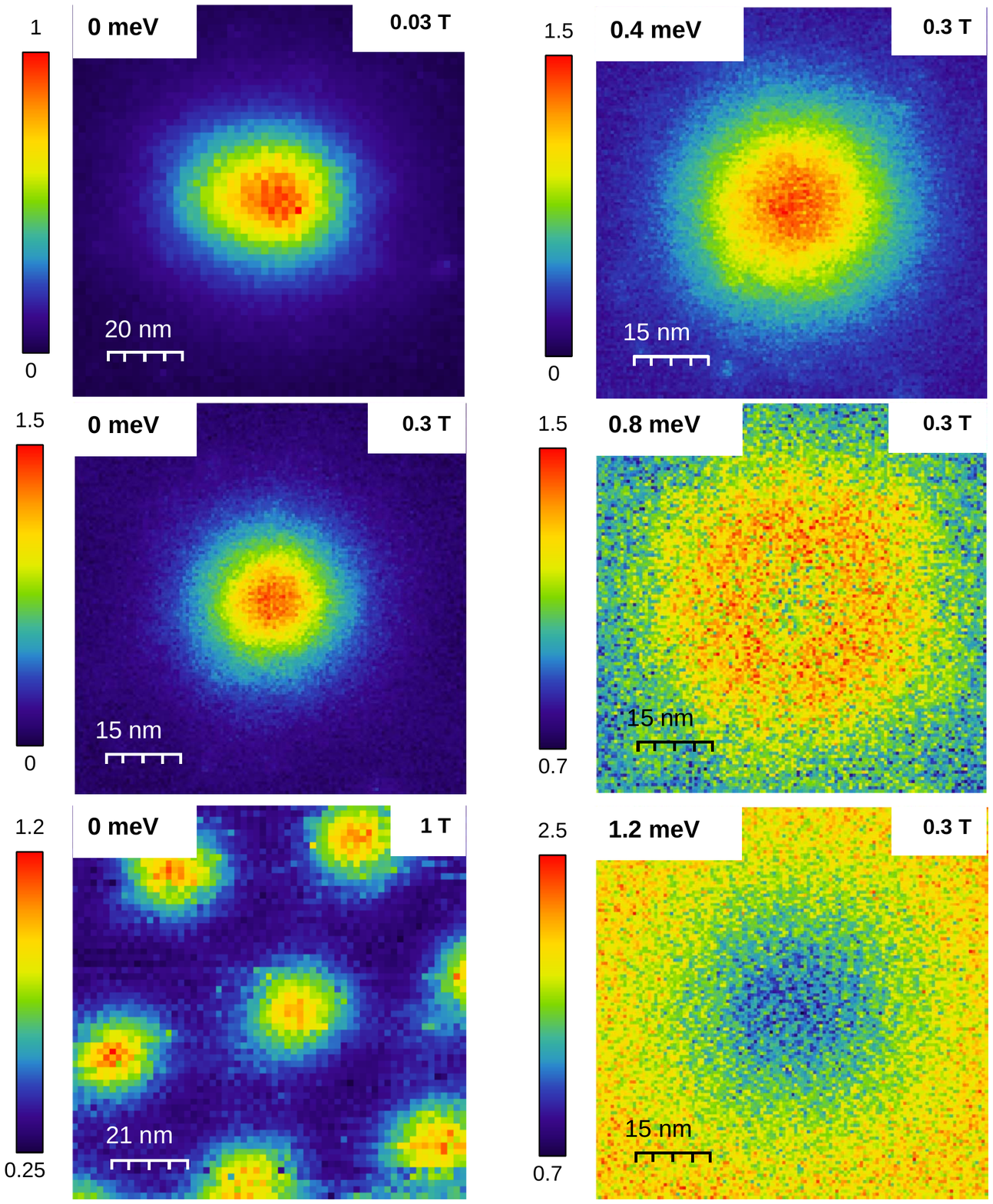}
\vskip -2 cm
\caption{Left panels: the zero-bias   conductance maps in 2H-NbSe$_{1.8}$S$_{0.2}$ at    $T=0.8\,$K  at   fields indicated. Right panels:  conductance maps of an isolated vortex for   bias voltages  indicated.}
\label{fa2}
\end{center}
\end{figure}

The experiment consists of home built low-noise dilution refrigerator STM system as described in Refs.\,\onlinecite{Suderow11}, \onlinecite{Galvis15}. The tunneling conductance curves are taken with an energy resolution of about 15 $\mu $eV \cite{Resolucion1,Resolucion2}. We use an Au tip, which we clean by repeated indentation onto the Au cleaning pad \cite{Rodrigo04}. The samples were cleaved in-situ  to obtain fresh surfaces.  Atomic resolution was   consistently achieved in all compounds discussed in this work. The magnetic field is applied perpendicular to the plate-like samples.
At the measurement temperature (of  0.15\,K for $\beta-$Bi$_2$Pd, 0.1\,K for 2H-NbS$_2$ and 0.8\,K for 2H-NbSe$_{1.8}$S$_{0.2}$) we can safely assume that the local conductance is proportional to the local  DOS, so that we can  replace the measured conductance $\sigma_0$ with the DOS $N$'s.

To make the fits shown in the publication (red lines of Fig.\,1), we use a single gap fit for $\beta-$Bi$_2$Pd and two gaps $\Delta_{1,2}$ with a Gaussian smearing of the DOS having a width of $\delta_{1,2}$ (see Table I).

\begin{table}[h]
\begin{tabular}{c | c | c }
Compound & $\Delta_{1,2} (meV) $ & $\delta_{1,2}$ (meV)\\
\hline
$\beta$-Bi$_2$Pd & 0.76, - & -,- \\
2H-NbSe$_{1.8}$S$_{0.2}$ & 0.78, 1 & 0.12, 0.12\\
2H-NbS$_2$ & 0.5, 1 & 0.16, 0.22\\
\end{tabular}
\caption{Superconducting parameters of the compounds studied used to obtain the red lines in Fig.\,1 of the publication.}
\end{table}

A few images of isolated vortices and of the vortex lattice are shown in   left panels of Fig.\ S1 for a set of magnetic fields.  The right panels show the evolution of the images with changing bias voltage at $H=0.3\,$T. Note that, contrary to the much discussed case of star-shaped vortices in 2H-NbSe$_2$, the vortices here are round due to scattering by the S impurities. As shown in the tunneling conductance curves, at different distances from the core center (Fig.\ S2), there is a zero bias peak. The zero bias peak is smeared due to scattering by the S substitution with respect to the pure compound as shown previously by  \cite{Renner}. Nevertheless, it is still clearly measurable.

We have also measured spatial distributions  $N(r,\epsilon)$ at finite bias energies $\epsilon$. In Fig.\,S3 we show the radially averaged $N(r,\epsilon)$ for all three compounds.
Note that the shape of the conductance vs bias voltage remains roughly the same in increasing     fields in $\beta-$Bi$_2$Pd, whereas for the other two compounds the conductance  tends to increase  away from the core with increasing $H$. This  tendency is seen by comparing the curves obtained from the zero bias data in Figs.\ 2 and 3 of the main text.

\begin{figure}[h]
\begin{center}
\includegraphics[width=9cm]{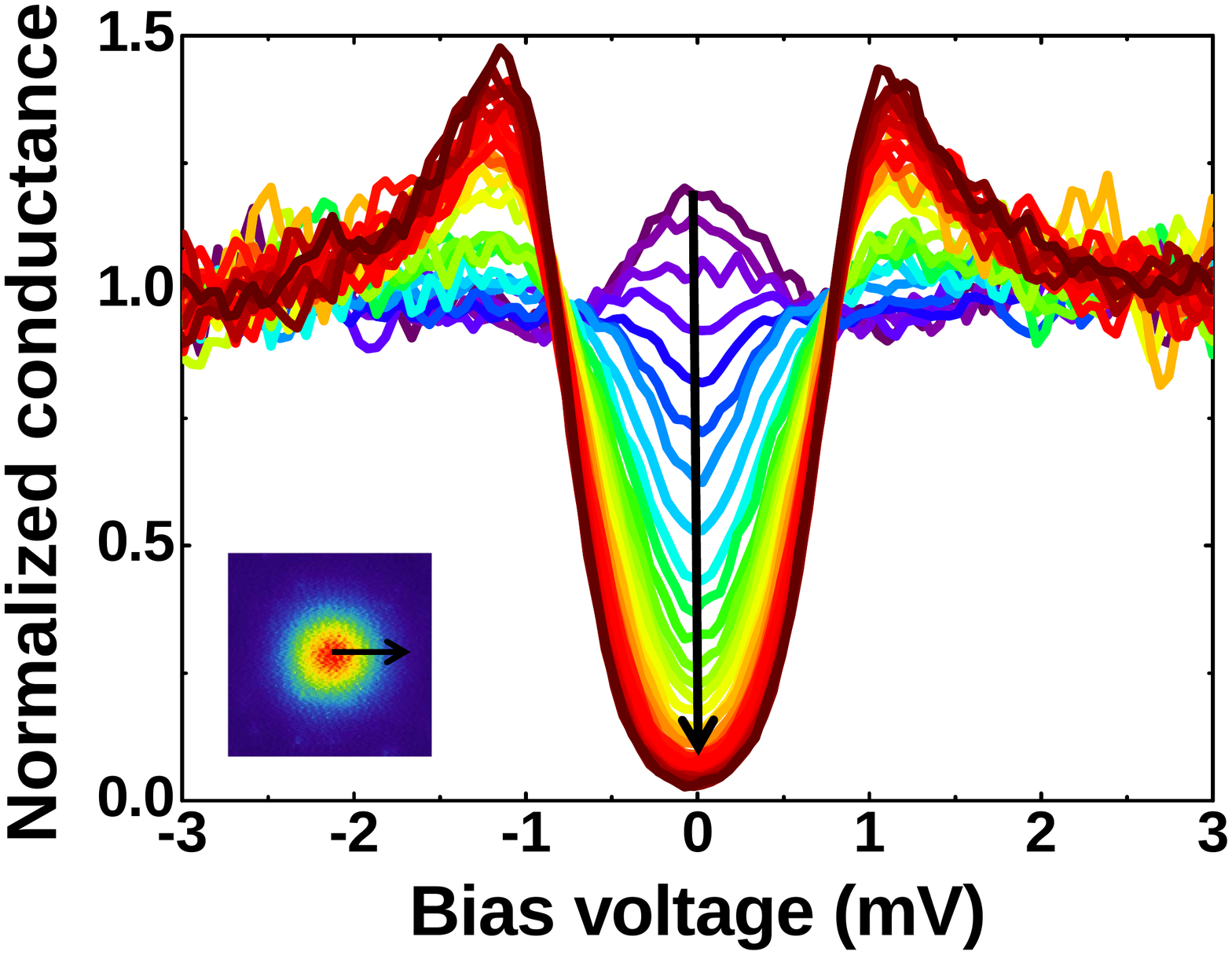}
\vskip -1 cm
\caption{Tunneling conductance vs position at $H= 0.02\,$T in 2H-NbSe$_{1.8}$S$_{0.2}$ along the line given by the black arrow in the inset, which shows the distribution of  zero bias conductance near the vortex.}
\label{fa2}
\end{center}
\end{figure}

\begin{figure}[h]
\begin{center}
\includegraphics[width=10cm]{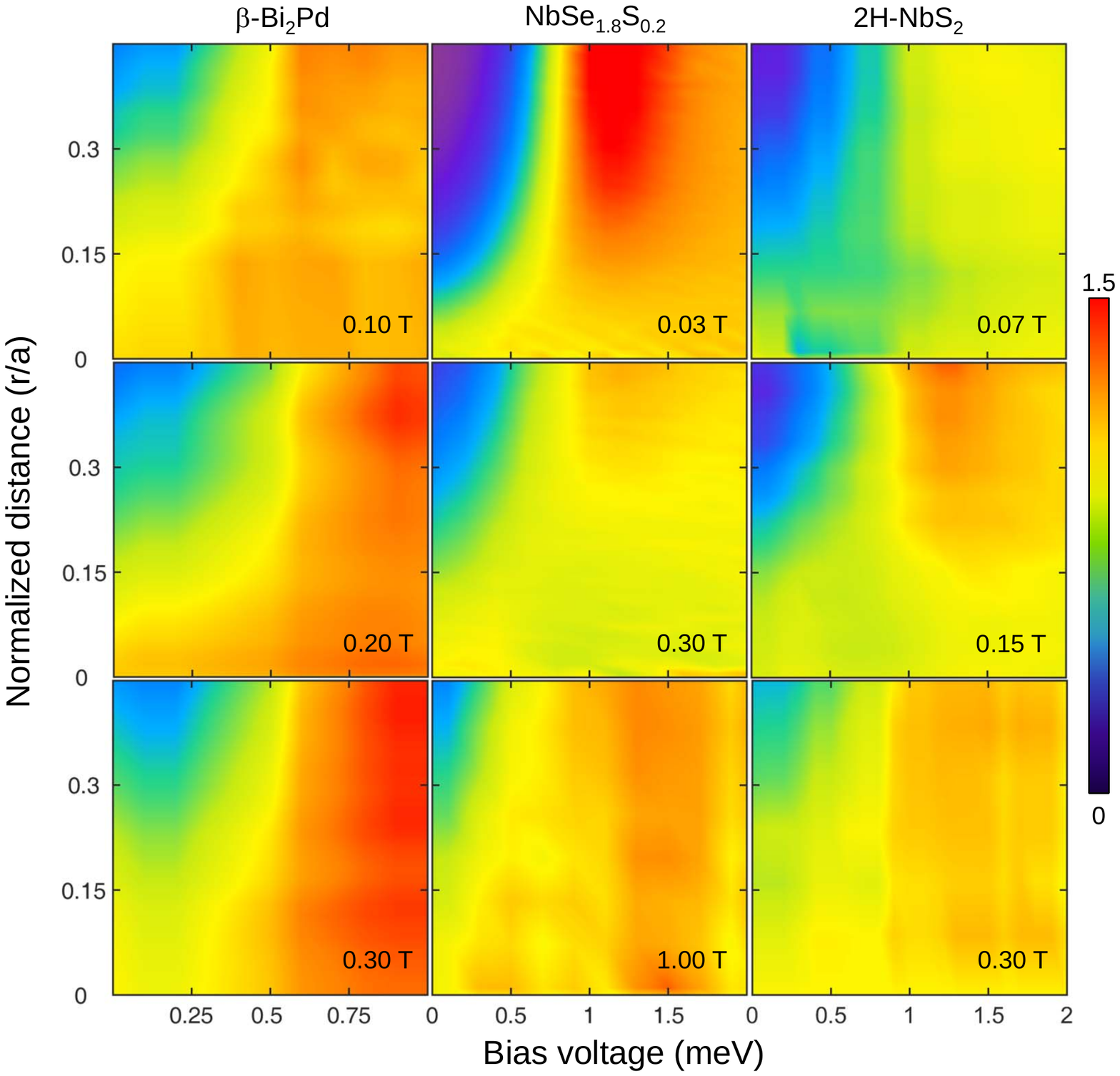}
\vskip -4.5 cm
\caption{The tunneling conductance normalized at its value at high bias voltages is radially averaged around a vortex center and plotted  vs distance $r/\xi_{c2}$ (with $\xi_{c2}$ being the coherence length obtained from $H_{c2}$) from the core (vertical axis) and the bias voltage (horizontal axis) for different magnetic fields.}
\label{f4}
\end{center}
\end{figure}

\subsection{On Caroli-deGennes-Matricon states in vortex cores}

The tunneling conductance within the vortex core shows zero bias peaks in 2H-NbSe$_{1.8}$S$_{0.2}$ and in 2H-NbS$_2$ due to Caroli-deGennes-Matricon core states. We do not take these peaks into account in our model. These provide a zero bias conductance slightly above one (see Fig.\,S2 for 2H-NbSe$_{1.8}$S$_{0.2}$ and Ref.\cite{NbS2} for 2H-NbS$_2$). The shape of the vortex core is not significantly influenced by these peaks. Note that, in the paper, we calculate 

\begin{equation}
\sigma =\frac{\sigma_0(r)-\sigma_0(r^*)}{\sigma_0(0)-\sigma_0(r^*)}\,.
\label{eq1}
\end{equation}

Thus, the magnetic field dependence of $\sigma_0(0)$ and of $\sigma_0(r^*)$ do not influence $\sigma$.

On the other hand, note that in previous work we have demonstrated that the absence of the core shape anisotropy in 2H-NbS$_2$ is due to the absence of a charge density wave (CDW) in this compound. Thus, CDW originates the in-plane anisotropy of the vortex core in 2H-NbSe$_2$ \cite{NbS2}.

Here, we observe the CDW in 2H-NbSe$_{1.8}$S$_{0.2}$ (Fig.\,S4). The loss of the in-plane anisotropy of the vortex core is here due to scattering by S impurities, as discussed previously in similar samples \cite{Renner}.

\begin{figure}[h]
\begin{center}
\includegraphics[width=9cm]{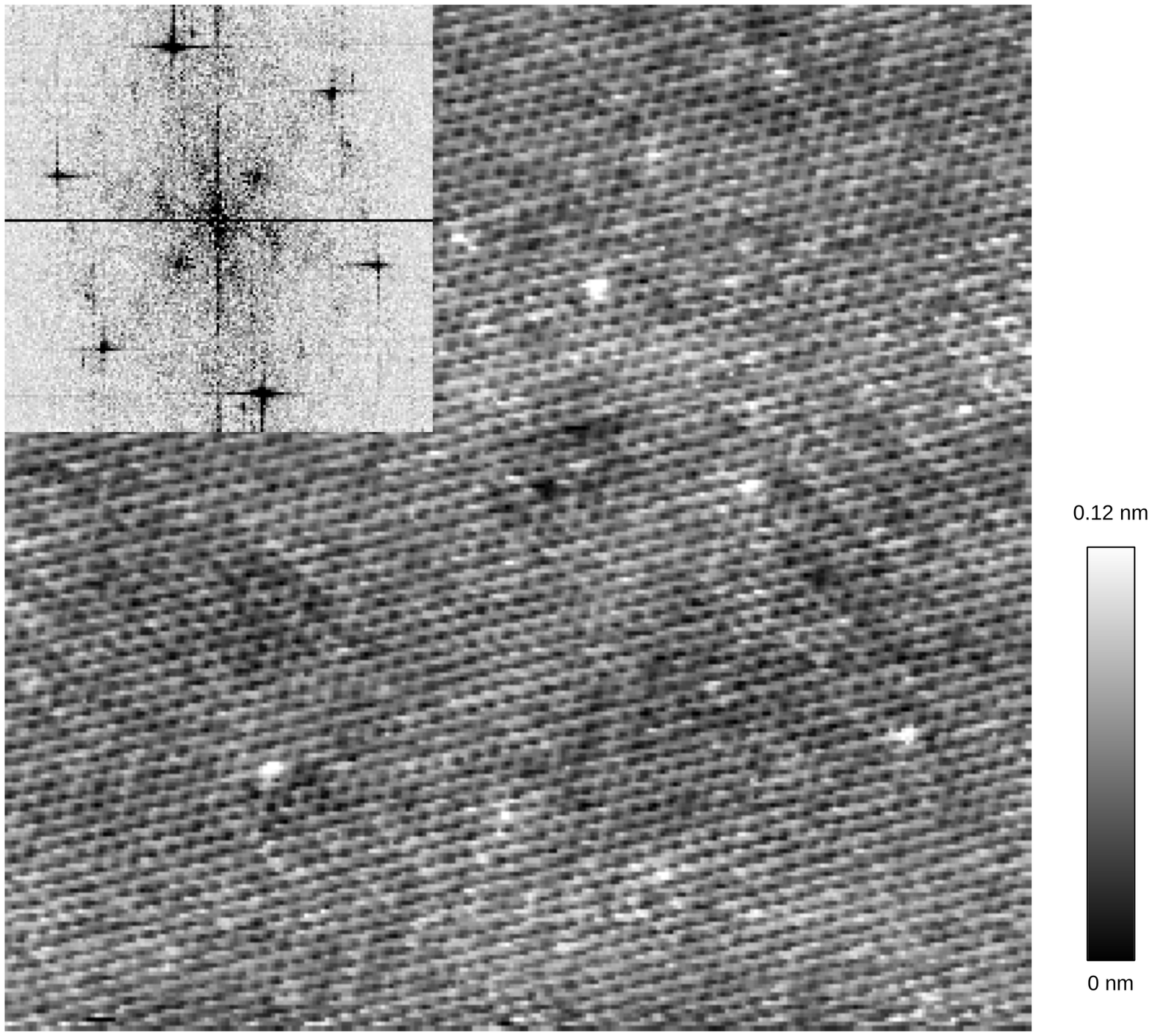}
\caption{Atomic resolution topography of 2H-NbSe$_{1.8}$S$_{0.2}$ showing the CDW in an area with a lateral size of 20 nm. The Fourier transform is shown in the upper left inset. The CDW wavevectors are indistinguishable, within experimental error, from the pure compound. The modulation is three-fold, with q=0.29 1/\AA\ for the atomic lattice and 0.085 1/\AA for the CDW, within an error of 20\% for the absolute values. The CDW is located at a distance of 34\% of the atomic lattice, i.e. $1/3$ within accuracy of the relative values, which we estimate to be around 5\% from the width of the Fourier transform peaks. Note also that there are sizeable variations in topographic contrast that are certainly due to scattering by S defects.}
\label{fa2}
\end{center}
\end{figure}

\subsection{Sample growth of 2H-NbSe$_{1.8}$S$_{0.2}$}

The synthesis of 2H-NbSe$_{1.8}$S$_{0.2}$ was performed in a typical solid state reaction. The elements were mixed in a stoichiometric ratio, sealed inside the evacuated quartz ampoule and heated from room temperature up to 900$^{\circ}$C at 1.5$^{\circ}$C/min. The sample  was kept at constant temperature during 14 days and then was slowly cooled down (0.07$^{\circ}$C/min). To obtain large single crystals, we mixed four mmol with I$_2$ as a transport agent ([I$_2]\approx$ 5 mg$/cm^3$) in  evacuated quartz tube, which was placed inside a three-zone furnace. We placed the material in the leftmost zone and heated the other two zones for three hours up to 700$^{\circ}$C and kept them at this temperature for one day. After that, the leftmost zone was heated to 750$^{\circ}$C within three hours and we established temperature gradients as 750$^{\circ}$C / 700$^{\circ}$C / 725$^{\circ}$C. These temperatures were kept for 22 days and then the oven was switched off for cooling.

The  crystals so formed were analyzed by inductively coupled plasma spectrometry and by powder X-ray diffraction (Fig.\ S5). The  elements content was Nb:36.9 $\pm 1.0$\%,Se:58.8 $\pm 1.5$ \% and S:2.4$\pm 0.2$ \% in good agreement with the expected values for 2H-NbSe$_{1.8}$S$_{0.2}$. Refinement of the X-ray pattern revealed a hexagonal lattice with a P63/mmc space group and a unit cell of $a = b = 3.4323(3)$ \AA, $c = 12.513(1)$ \AA, $\alpha$ = $\beta$ = 90$^{\circ}$ and $\gamma$ = 120$^{\circ}$. These results are only slightly different from those for   pure  2H-NbSe$_2$ ($a = b = 3.4425(5)$ \AA, $c = 12.547(3)$ \AA, $\alpha$ = $\beta$ = 90$^{\circ}$ and $\gamma$ = 120$^{\circ}$) \cite{Meerschaut}.

\begin{figure}[h]
\begin{center}
\includegraphics[width=9cm]{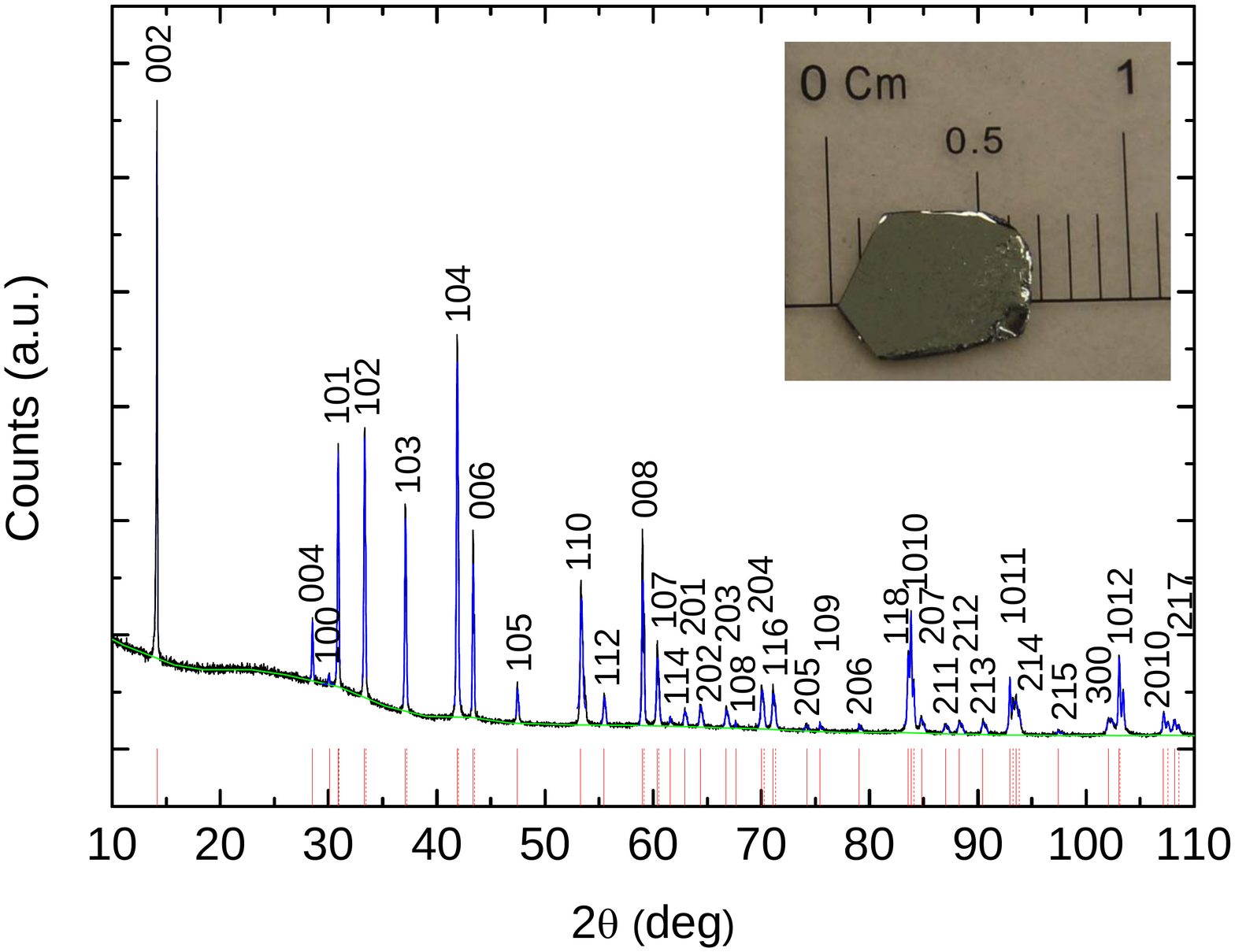}
\caption{XRPD experimental pattern of 2H-NbSe$_{1.8}$S$_{0.2}$ single crystal (black) and corresponding fit (peaks in blue and background in green). Inset: photography of a single crystal. The fit gives: a = b = 3.4323(3) \AA\ and c = 12.513(1) \AA, hexagonal crystal system with P63/mmc space group, $X^2$ = 2.22·10$^{-5}$ and a Snyder's figure of merit of 44.3158.}
\label{fa1}
\end{center}
\end{figure}

\subsection{Resistivity for $\beta$-Bi$_2$Pd, 2H-NbSe$_{1.8}$S$_{0.2}$ and 2H-NbS$_2$}

In the Fig.\,S6 we provide the temperature dependence of the resistivities normalized to the ambient temperature value for all three compounds studied in the publication.

\begin{figure}[h]
\begin{center}
\includegraphics[width=9cm]{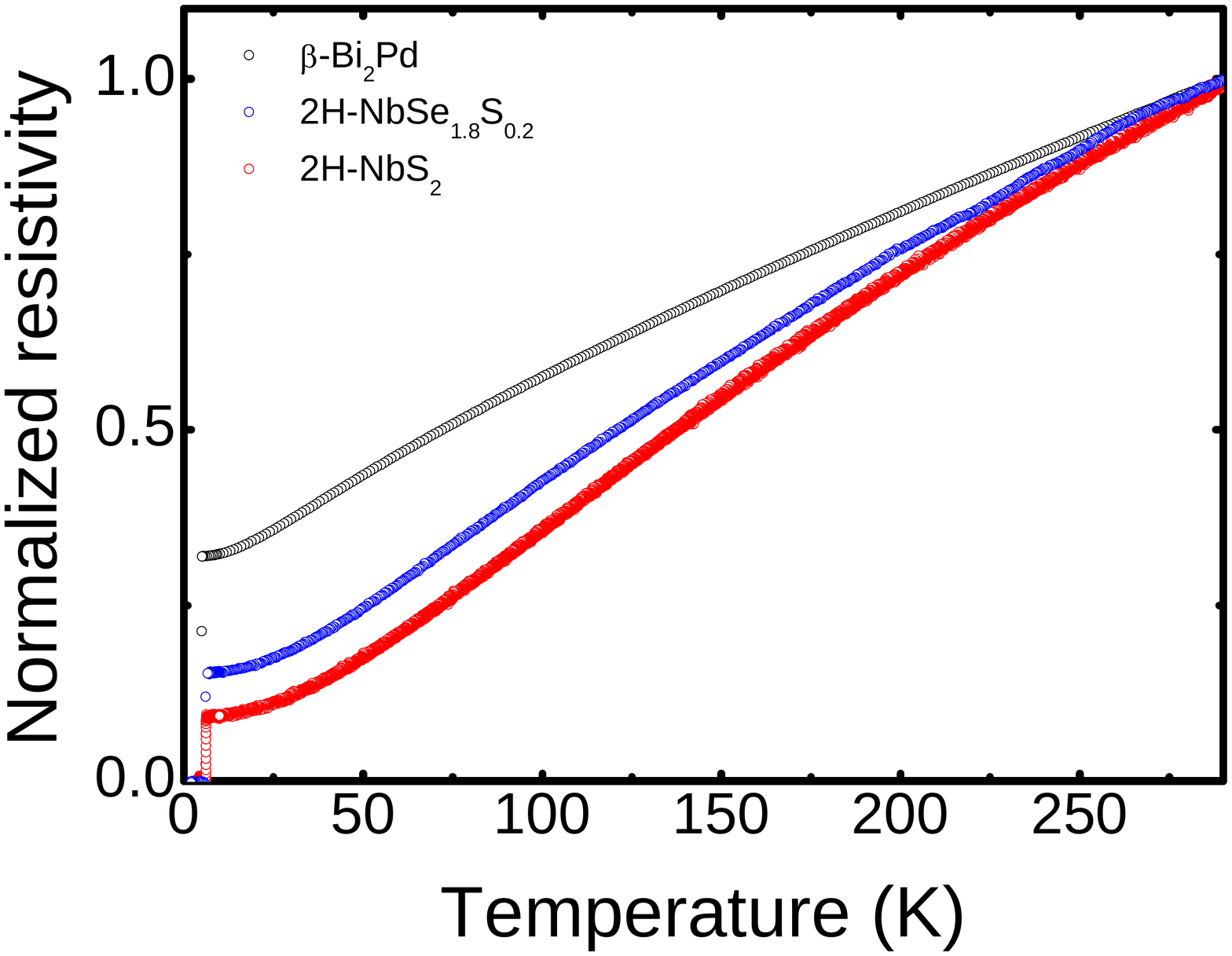}
\caption{Temperature dependence of the resistivity for $\beta$-Bi$_2$Pd, 2H-NbSe$_{1.8}$S$_{0.2}$ and 2H-NbS$_2$ normalized to the ambient temperature value.}
\label{fa2}
\end{center}
\end{figure}

\end{document}